\documentclass[letter,11pt]{article}
\usepackage[top=1in,bottom=1in,left=1in,right=1in]{geometry}
\usepackage{slashed}
\usepackage{epsfig}
\usepackage{comment}
\usepackage{cancel}
\usepackage{bbm}
\usepackage{array}
\usepackage{bigints}
\usepackage{booktabs}
\usepackage{color}
\usepackage{dsfont}
\usepackage{float}
\usepackage{framed}
\usepackage{graphicx}
\usepackage{indentfirst}
\usepackage{mathrsfs}
\usepackage{multirow}
\usepackage{subdepth}
\usepackage{titlesec}
\usepackage[dotinlabels]{titletoc}
\usepackage{wrapfig}
\usepackage[all]{xy}
\usepackage[vcentermath]{youngtab}
\usepackage{relsize}
\usepackage{hyperref}
\numberwithin{equation}{section}

\usepackage[utf8]{inputenc}
\usepackage{slashed}
\usepackage{amsmath}
\usepackage{amsfonts}
\usepackage{amssymb}
\usepackage{cite}
\usepackage{mathtools}

\usepackage{cleveref}
\crefname{section}{§}{§§}
\Crefname{section}{§}{§§}

\def\ip{${\mathscr I}^+$}

 \def\p{\partial}
 \def\bz{{\bar z}}
 \def\bw{{\bar w}}
 
\def\0{{(0)}}
\def\1{{(1)}}
\def\2{{(2)}}

\def\ci{{\mathscr I}}

\def\<{\langle }
\def\>{\rangle }
\def\bw{{\bar w}}

\newcommand{\bea}{\begin{eqnarray}}
\newcommand{\eea}{\end{eqnarray}}
\newcommand{\be}{\begin{equation}}
\newcommand{\ee}{\end{equation}}
\newcommand{\ba}{\begin{align}}
\newcommand{\ea}{\end{align}}

   \makeatletter
  \let\over=\@@over \let\overwithdelims=\@@overwithdelims
  \let\atop=\@@atop \let\atopwithdelims=\@@atopwithdelims
  \let\above=\@@above \let\abovewithdelims=\@@abovewithdelims
\renewcommand\section{\@startsection {section}{1}{\z@}%
                                   {-3.5ex \@plus -1ex \@minus -.2ex}
                                   {2.3ex \@plus.2ex}%
                                   {\normalfont\large\bfseries}}

\renewcommand\subsection{\@startsection{subsection}{2}{\z@}%
                                     {-3.25ex\@plus -1ex \@minus -.2ex}%
                                     {1.5ex \@plus .2ex}%
                                     {\normalfont\bfseries}}

\newcommand{\beq}{\begin{equation}}
\newcommand{\eeq}{\end{equation}}
\newcommand{\beqa}{\begin{eqnarray}}
\newcommand{\eeqa}{\end{eqnarray}}
\newcommand{\beqar}{\begin{eqnarray*}}

\def\[{\big[}
\def\]{\big]}

\def\bz{{\bar z}}

\def\be{{\bar \epsilon}}

\def\bw{{\bar w}}

\def\CA{{\mathcal A}}

\def\CF{{\mathcal F}}

\def\CI{{\mathcal I}}
\def\ci{{\mathcal I}}

\def\CO{{\mathcal O}}
\def\CP{{\mathcal P}}

\def\+{{(+)}}
\def\-{{(-)}}
\def\0{{(0)}}
\def\1{{(1)}}
\def\2{{(2)}}
\def\3{{(3)}}

\def\ip{${\cal I}^+$}

\def\be{\begin{equation}}
\def\ee{\end{equation}}

\newcommand{\bd}[1]{\begin{fmffile}{#1}\begin{fmfgraph*}}
\newcommand{\ed}{\end{fmfgraph*}\end{fmffile}}

\begin{document}
\begin{titlepage}
\unitlength = 1mm
\ \\
\vskip 3cm
\begin{center}

{\LARGE{\textsc{Color Memory }}}

\vspace{0.8cm}
Monica Pate, Ana-Maria Raclariu and Andrew Strominger

\vspace{1cm}

{\it  Center for the Fundamental Laws of Nature, Harvard University,\\
Cambridge, MA 02138, USA}

\vspace{0.8cm}

\begin{abstract}
A transient color flux across null infinity in classical Yang-Mills theory is considered.
It is shown that  a pair of test  `quarks' initially in a color singlet generically acquire net color as a result of the flux. 
A nonlinear formula is derived for the relative color rotation of the quarks. For weak color flux the formula  linearizes to the Fourier transform of the soft gluon theorem. 
This color memory effect is the Yang-Mills analog of the gravitational memory effect. 
\end{abstract}

\vspace{1.0cm}

\end{center}

\end{titlepage}

\pagestyle{empty}
\pagestyle{plain}

\def\vx{{\vec x}}
\def\p{\partial}
\def\po{$\cal P_O$}
\def\i{{\rm initial}}
\def\f{{\rm final}}

\pagenumbering{arabic}
 

\tableofcontents

\section{Introduction}
The gravitational memory effect in general relativity \cite{Zeldovich1974, BraginskiiThorne1987,Christodoulou:1991cr} concerns  a subtle and beautiful aspect of the behavior of inertial detectors in the weak field region far from  gravitating sources. 
It is a key observational link interconnecting  asymptotic  symmetries and soft theorems \cite{Strominger:2014pwa}. In this paper we derive the analog of this effect in classical nonabelian gauge theory. The memory effect in abelian gauge theory was discussed in \cite{Bieri:2013hqa,Pasterski:2015zua,Susskind:2015hpa}.

	The color memory effect can be seen by  two test `quarks'  in a color singlet stationed at  fixed large radius near future null infinity \ip  and 
 fixed angles $\Theta_ \alpha$ where the label $\alpha = 1,2$ distinguishes the two quarks.   In order for the statement that they are in a 
	color singlet to have any meaning, we must specify a flat connection  $A=i UdU^{-1}$ on \ip.  Here $U$ is an element in the gauge group $G$. For simplicity we take the initial value at retarded time $u_i$ to be 
	\be U(u_i)=1,\ee although the generalization is straightforward. Now consider the effect of color flux or radiation through \ip, which we take to begin after the initial time $u_i$ and end before some final time $u_f$.  Over this time interval, while  the position of 
	each quark is pinned to a fixed radius and angle, the color of each quark $Q_\alpha$  evolves
	according to 
	\begin{equation} \p_u Q_\alpha= i A_u(\Theta_\alpha) Q_\alpha, \end{equation}
	where $A_u$ denotes a component of the gauge field near \ip. 
	It is convenient to use temporal  gauge 
 	\be \label{tempgauge} A_u=0,\ee
	so that the quarks do not change their colors.

By assumption at late times $u>u_f$ the field strength vanishes and the connection is again flat. However, we will see that the classical constraint equation on \ip\ implies that generically \be U(u_f)\neq 1.\ee
This is color memory: the connection `remembers' some aspects of the color flux. 
This means that our two initially color-singlet quarks will no longer be in a color singlet after the passage of the color flux. Parallel transport from $\Theta_1$ to $\Theta_2$ to compare their colors will reveal a relative color rotation between the quarks  \be\label{rft} U(u_f,\Theta_2 )U^{-1}(u_f,\Theta_1).\ee
This conclusion does not depend on the temporal gauge choice.
The main result of this paper is a nonlinear  formula for $U(u_f)$ in terms of the color flux through \ip.

Classical vacua in nonabelian gauge theory are degenerate and labelled by the flat connections on the `celestial sphere' at \ip \cite{He:2015zea}. These vacua are related by the action of spontaneously broken `large' gauge symmetries that do not die off at infinity. Color flux through \ip\ is a domain wall which induces transitions between the degenerate vacua. The color memory effect measures these transitions. If the celestial sphere is initially tiled with test quarks pointing in the same direction in color space, after the passage of the color flux they will point in different directions upon parallel transport by \eqref{rft}. This enables one to  read off the difference   between the initial and final flat connections.

Gravity and abelian gauge theory essentially become free theories at \ip. 
In these cases the memory effect is a linear function of the appropriate flux at \ip\ and is given by the Fourier transform of the soft theorem. 
This is not the case for nonabelian gauge theory, for which nonlinear effects, albeit in substantially weakened form, persist all the way to the boundary of Minkowski space. 
Finding the finite classical memory effect requires solving an interesting  nonlinear PDE on the sphere. 
We present and solve this equation to the first two orders in weak field perturbation theory,\footnote{It would be interesting to either find closed form solutions or prove that they exist.} with the first order given by the Fourier-transformed soft gluon theorem. Interestingly, the full nonlinear PDE 
 has appeared previously in the QCD literature. See for example work of McLerran and collaborators \cite{McLerran:1993ka} who were studying gluon distribution functions inside hadrons at small Bjorken scale $x$.
This suggests that the color memory effect has already been encountered in some form, just not 
called by that name or related to the vacuum degeneracy of flat connections. This relation merits further investigation.

Efforts to measure the gravitational memory effect are underway at LIGO \cite{Lasky:2016knh} and the pulsar timing array\cite{vanHaasteren:2009fy,Wang:2014zls}. Measurement of $SU(3)$ color memory is of course difficult because of confinement.  The measurement must take place on an energy scale above the confinement scale yet below that of the dynamical process. Indeed, as the basic equation of color memory has been previously  encountered \cite{McLerran:1993ka}, color memory may already have been measured!  We leave possible experimental applications of color memory  to future studies.

\section{Preliminaries}

In this section we present  notations, conventions and an asymptotic expansion of the field equations. 

We consider a nonabelian gauge theory with gauge group $G$ and elements $g_R$ in representation $R$. Hermitian generators $T^a_R$ in  the representation $R$ obey 
\begin{equation}
[T^a_R, T^b_R] = i f^{abc} T^c_R,
\end{equation}
where the $a $ runs over the dimension of the group and the sum over repeated indices is implied.  
We denote the $4$-dimensional gauge potential $\CA_{\mu} = \CA_{\mu}^a T^a_R$  with spacetime index $\mu=0,1,2,3$. 

Since we will be interested in the asymptotic expansions of fields near future null infinity ($\CI^+$), it is convenient to introduce retarded coordinates in which the Minkowski metric reads 
\begin{equation}
ds^2 = -du^2 - 2 du dr + 2r^2\gamma_{z\bar{z}} dz d\bar{z},
\end{equation}
where $(z,\bz)$ are stereographic coordinates on the celestial sphere with  $\gamma_{z\bz} = \frac{2}{(1 + z\bar{z})^2}$ the unit round metric.
The equations of motion are 
\begin{equation}
\label{YM}
\nabla^{\nu} \CF_{\nu\mu} - i[\CA^{\nu}, \CF_{\nu\mu}] = g_{YM}^2 j_{\mu}^M ,
\end{equation}
where $j_{\mu}^M$ is the matter color current, $g_{YM}$ the gauge coupling  and the field strength is\begin{equation}
\CF_{\mu\nu} = \p_{\mu}\CA_{\nu} - \p_{\nu} \CA_{\mu} - i[\CA_{\mu}, \CA_{\nu}].
\end{equation}
The theory is invariant under the gauge transformations
\begin{equation}
\begin{split}
\CA_{\mu} &\rightarrow  g_R \CA_{\mu} g_R^{-1} + ig_R\partial_{\mu} g_R^{-1},\\
j_{\mu}^M &\rightarrow  g_R j_{\mu}^M g_R^{-1}.
\end{split}
\end{equation}

Working in temporal gauge \eqref{tempgauge},   we expand the remaining components of the gauge field near \ip in inverse powers of $r$ \cite{Strominger:2017zoo}
\begin{equation} 
\CA_r(u,r,z, \bz) = \frac{1}{r^2}A_r(u,z,\bz) + \mathcal{O}(r^{-3}),\qquad
\CA_z (u,r,z ,\bz) = A_z(u,z,\bz) + \mathcal{O}(r^{-1}).
\end{equation}
These fall-off conditions ensure finite charge and energy flux through $\CI^+$ and they are preserved by large gauge transformations that approach $(z, \bz)$-dependent Lie group-valued functions on the celestial sphere\cite{Strominger:2013lka,He:2015zea}.
The leading behavior of the field strength is then
\begin{equation}
\CF_{ur} = \frac{1}{r^2}F_{ur} + \CO(r^{-3}), \qquad \CF_{uz} = F_{uz} + \CO(r^{-1}), \qquad \CF_{z\bz} = F_{z\bz} + \CO(r^{-1}),
\end{equation}
where 
\begin{equation}
F_{ur} =\p_uA_r , \qquad F_{uz}=\p_uA_z , \qquad F_{z\bz} =\p_zA_\bz-\p_\bz A_z-i[A_z,A_\bz].
\end{equation} 
In retarded coordinates, the $u$ component of \eqref{YM} reads
\begin{equation}
\nabla^r \CF_{ru} + \nabla^A \CF_{Au} - i\left([\CA^r, \CF_{ru}] + [\CA^A, \CF_{Au}] \right) = g_{YM}^2 j_u^M,
\end{equation}
where here and hereafter $A, B,...$ run over the $S^2$ coordinates $(z, \bz)$.
At leading order in the large-$r$ expansion we find
\begin{equation}
\label{YM-comp}
-\partial_u F_{ru} + D^A F_{Au} = J_u ,
\end{equation}
  where $D_A$ is the covariant derivative on the unit $S^2$ and its indices are raised and lowered with $\gamma_{AB}$. The asymptotic color flux  is
  \be J_u= i\gamma^{AB}[A_B, F_{Au}] +g^2_{YM} \lim_{r \to \infty }   \left [r^2 j_u^M \right ]. \ee
This includes a quadratic term from the gauge potential itself, as gluons contribute to the color flux. Note that the left hand side of 
\eqref{YM-comp}  is linear in the gauge potential. 
\section{Color memory effect}
We wish to compute  the change in the vacuum, or flat connection, induced by color flux $J_u$ through $\ci^+$.  For simplicity we consider configurations with no color flux or magnetic fields  prior to some initial retarded time $u_i$ and  after some final retarded time $u_f$:
\begin{equation}
\label{Fzbz}
 F_{z\bz}|_{u < u_i} = F_{uz}|_{u < u_i} = F_{z\bz}|_{u > u_f} = F_{uz}|_{u > u_f}= 0,
\end{equation}
and where the Coulombic component of the field strength is constant over the sphere at initial and final times
\be
	F_{ru} (u_i, z, \bz)  = F_{ru} (u_i) , \quad \quad \quad F_{ru} (u_f, z, \bz)  = F_{ru} (u_f).
\ee
The leading component of $F_{z\bz}$, unlike those of $F_{ru}$ or $F_{uz}$, does not linearize at \ip. This is the source of the nonlinearity of color memory.  
In this case, \eqref{YM} determines $A_z|_{u > u_f}$ in terms of the color flux $J_u$,  $A_z|_{u < u_i}$ and the  initial and final electric fields $F_{ru}|_{u < u_i}$ and $F_{ru}|_{u > u_f}$ (which may be set to zero in some applications). By \eqref{Fzbz}, the transverse components of the boundary gauge fields are pure gauge and hence related by a large gauge transformation. The large gauge transformation determining the change in $A_z$ across $\CI^+$ can be found by solving \eqref{YM-comp} subject to the boundary conditions \eqref{Fzbz}.

We now determine the change of $A_{z}$ across $\mathcal{I}^+$. Integrating \eqref{YM-comp} over $u$, we find
\begin{equation}
\label{intYM}
-D^A \Delta A_{A} = \int_{u_i}^{u_f} du  J_u + \Delta F_{ru},
\end{equation}
where 
\begin{equation}
 \Delta A_z = A_z(u_f,z,\bz) - A_z(u_i, z, \bz), \qquad \Delta F_{ru} = F_{ru}(u_f ) - F_{ru}(u_i).
\end{equation}
Let us  set $A_A(u_i) = 0$ by performing a large gauge transformation  and define
\begin{equation}
J_{z\bz} = -\gamma_{z\bz} \left (\int_{u_i}^{u_f} du J_u+ \Delta F_{ru} \right ) .
\end{equation} 
Then, \eqref{intYM} reduces to 
\begin{equation}
\label{div}
\p_z A_{\bz}(u_f) + \p_{\bz} A_z(u_f) = J_{z\bz}.
\end{equation}
 
The general solution to the boundary conditions \eqref{Fzbz} is
\beq  \label{flatA}
	A_z (u_f) = i U \p_zU^{-1} ,
\eeq
where $U = U(z, \bz) \in G$.  Substituting this solution in \eqref{div}, we obtain an equation for the large gauge transformation $U$ relating initial and final flat connections on $S^2$
\beq
	i \p_z  \left (U \p_\bz U^{-1}  \right)  +i \p_\bz  \left (U \p_z U^{-1}  \right)  = J_{z\bz}.  
\eeq
The color memory effect is defined by the solution to this equation.  $U(z,\bz)$ determines  the flat connection on $S^2$ after the color flux passes through \ip.

To solve this equation perturbatively in $J_{z\bz}$, we first invert \eqref{flatA}
\beq \label{UfromA}
	U(z, \bz) = \CP \left \{ \exp \left (i \int^{(z, \bz)}_{(0,0)} dw A_w + d \bw A_{\bw}\right)\right \},
\eeq
where $\CP$ denotes path-ordering. $U(z,\bz)$ is  independent of path because  $A$ is flat. 
Then, since the asymptotic gauge potential $A(u_f)$ is a 1-form on $S^2$, it can  be parametrized as
\begin{equation}
A_z(u_f) = \p_z \left(\alpha + i\beta \right), \qquad A_{\bz}(u_f) = \p_{\bz}\left(\alpha - i\beta \right),
\end{equation}
where $\alpha, \beta$ are Lie algebra-valued real functions on $S^2$. Substituting into \eqref{div}, we find
\begin{equation}
2\p_z\p_{\bz}\alpha = J_{z\bz},
\end{equation}
which is solved by
\begin{equation}
\alpha(z) = \frac{1}{4\pi}\int d^2w ~G(z, w)J_{w\bw}, \qquad G(z, w) = \log \frac{|z - w|^2}{(1+ z \bz) (1+ w \bw)}.
\end{equation}
The boundary conditions \eqref{Fzbz} give an additional differential equation for $\beta$ 
\begin{equation}
2\p_z\p_{\bz}\beta + [\p_z\alpha, \p_{\bz}\alpha] + [\p_z\beta, \p_{\bz}\beta] - i[\p_z\alpha, \p_{\bz}\beta] + i[\p_z\beta, \p_{\bz}\alpha] = 0.
\end{equation} 
This equation can be solved perturbatively  in $J_{z \bz}$ for $\beta$ which is given to leading order by
\beq
	\beta(z) =- \frac{1}{(4 \pi)^3 } \int d^2w d^2w' d^2 w'' ~G(z,w) \p_w G(w,w') \p_\bw G(w,w'') \left [J_{w' \bw'}, J_{w''\bw''} \right]  + \CO (J^3).
\eeq
 
To this order in perturbation theory, the expression for the gauge field   is 
\beq
	\begin{split}
	A_z &= \int \frac{d^2w}{4 \pi } \p_z G(z, w) \left [J_{w\bw} -i \int   \frac{d^2w'}{4 \pi} \frac{d^2 w''}{4 \pi} ~  \p_w G(w,w') \p_\bw G(w,w'') \left [J_{w' \bw'}, J_{w''\bw''} \right]  \right]+ \CO (J^3),\\
		A_\bz &= \int \frac{d^2w}{4 \pi } \p_\bz G(z, w) \left [J_{w\bw} +i \int   \frac{d^2w'}{4 \pi} \frac{d^2 w''}{4 \pi} ~  \p_w G(w,w') \p_\bw G(w,w'') \left [J_{w' \bw'}, J_{w''\bw''} \right]  \right]+ \CO (J^3),
	\end{split}
\eeq
from which the large gauge transformation associated to the vacuum transition can be directly obtained via \eqref{UfromA}. It may be seen that the first term is the Fourier transform of the soft gluon theorem, while the second is the leading nonlinear correction for finite color flux.

\section*{Acknowledgements}
 This work was supported in part by DOE grant DE-SC0007870.

\end{document}